\documentclass[aps,prb,showpacs,twocolumn,superscriptaddress]{revtex4} 

\usepackage[dvipdfmx]{graphicx}
\usepackage[]{graphicx,color}% Include figure files
\usepackage[version=3]{mhchem}
\usepackage{bm} 
\usepackage{physics} 
\usepackage{ulem}

\begin{document}

%Title of paper
\title{Evolutionary search for superconducting phases in the lanthanum-nitrogen-hydrogen system with universal neural network potential}

% repeat the \author .. \affiliation  etc. as needed
% \email, \thanks, \homepage, \altaffiliation all apply to the current
% author. Explanatory text should go in the []'s, actual e-mail
% address or url should go in the {}'s for \email and \homepage.
% Please use the appropriate macro foreach each type of information

% \affiliation command applies to all authors since the last
% \affiliation command. The \affiliation command should follow the
% other information
% \affiliation can be followed by \email, \homepage, \thanks as well.

\author{Takahiro Ishikawa}%
 \email{takahiro.ishikawa@phys.s.u-tokyo.ac.jp}
 \affiliation{%
 Department of Physics, The University of Tokyo, 7-3-1 Hongo, Bunkyo-ku, Tokyo 113-0033, Japan
 }%
 %\affiliation{%
 %ESICMM, National Institute for Materials Science, 1-2-1 Sengen, Tsukuba, Ibaraki 305-0047, Japan
 %}%
\author{Yuta Tanaka}%
 \affiliation{%
 Central Technical Research Laboratory, ENEOS Corporation, 8, Chidoricho, Naka-ku, Yokohama, Kanagawa 231-0815, Japan
  }%
\author{Shinji Tsuneyuki}%
 \affiliation{%
 Department of Physics, The University of Tokyo, 7-3-1 Hongo, Bunkyo-ku, Tokyo 113-0033, Japan
  }%
   
\date{\today}% It is always \today, today,
             %  but any date may be explicitly specified
%\date{September 10, 2005}

\begin{abstract}
%The maximum length is 600 characters including spaces.
Recently, Grockowiak \textit{et al.} reported ``hot superconductivity'' in ternary or multinary compounds based on lanthanum hydride [A. D. Grockowiak \textit{et al.}, Front. Electron. Mater. \textbf{2}, 837651 (2022)]. 
In this paper, we explored thermodynamically stable phases and superconducting phases in the lanthanum-nitrogen-hydrogen system (\ce{La_{$x$}N_{$y$}H_{$1-x-y$}}, $0 \leq x \leq 1$, $0 \leq y \leq 1$) at pressure of 20\,GP. We rapidly and accurately constructed the formation-enthalpy convex hull using an evolutionary construction scheme based on density functional theory calculations, extracting the candidates for stable and moderately metastable compounds by the universal neural network potential calculations. The convex hull diagram shows that more than fifty compounds emerge as stable and moderately metastable phases in the region of $\Delta H \leq 4.4$\,mRy/atom. In particular, the compounds are concentrated on the line of $x = 0.5$ connecting between LaH and LaN. We found that the superconductivity is gradually enhanced due to N doping for LaH and the superconducting critical temperature $T_{\rm c}$ reaches 8.77\,K in La$_2$NH with $y = 0.25$. In addition, we predicted that metastable La$_2$NH$_2$ shows the highest $T_{\rm c}$ value, 14.41\,K, of all the ternary compounds predicted in this study. These results suggest that it is difficult to obtain the hot superconductivity in the La-H compounds with N at 20\,GPa. 
\end{abstract}

% insert suggested PACS numbers in braces on next line
%\pacs{61.50.Ah, 61.66.Bi, 71.20.-b, 71.20.Be, 73.43.-f}
%\pacs{61.50.Ah, 61.66.Bi, 71.20.−b, 71.20.Be, 73.43.−f}
% insert suggested keywords - APS authors don't need to do this
%\keywords{}

%\maketitle must follow title, authors, abstract, \pacs, and \keywords
\maketitle

% body of paper here - Use proper section commands
% References should be done using the \cite, \ref, and \label commands
%\section{}
% Put \label in argument of \section for cross-referencing
%\section{\label{}}
%\subsection{}
%\subsubsection{}

\section{Introduction}

Metallic hydrides have attracted much attention as potential candidates for room-temperature superconductor~\cite{Ashcroft1968,Ashcroft2004}. 
In 2015, Drozdov \textit{et al.} performed electric resistivity measurements on hydrogen sulfide in diamond anvil cell to verify high-temperature superconductivity predicted in sulfur hydrides under high pressure~\cite{Li2014,Duan2014-SciRep}, and observed the superconductivity at 203\,K at pressure of 155\,GPa~\cite{Drozdov2015}. This discovery has led to further experimental and theoretical studies on stoichiometry, crystal structure, and superconductivity in sulfur hydrides under high pressure~\cite{Duan2015,Errea2015,Akashi2015,Errea2016,Ishikawa2016-SciRep,Akashi2016}, and x-ray diffraction measurements have clarified that sulfur hydride takes cubic H$_3$S in the high-temperature superconducting phase~\cite{Einaga2016}.  
In 2018 and 2019, Somayazulu \textit{et al.} and Drozdov \textit{et al.} observed that lanthanum hydride (LaH$_{10}$) shows the superconductivity at 250--260\,K in pressure region of 170--190\,GPa~\cite{Somayazulu2019,Drozdov2019-LaHx}, which exceeds the superconducting critical temperature ($T_{\rm c}$) in H$_3$S and is close to room temperature. 
In addition to these hydrides, first-principles calculations have predicted the superconductivity in binary hydrides under high pressure with respect to more than 60 elements~\cite{Ishikawa2019}, in which YH$_{10}$~\cite{Liu2017-LaH10}, MgH$_{6}$~\cite{Feng2015}, CaH$_{12}$~\cite{Semenok2020}, and AcH$_{10}$~\cite{Semenok2018} are predicted to show the $T_{\rm c}$ values over 200\,K. 

As the next stage for the exploration, recently, ternary hydrides are gathering attention as the candidate for higher $T_{\rm c}$ superconductivity. In 2019, Sun \textit{et al.} predicted from first-principles calculations that \ce{Li_{2}MgH_{16}} shows $T_{\rm c}$ of 475\,K at 250\,GPa~\cite{Sun2019}. In 2020, Sun \textit{et al.} predicted that the insertion of CH$_{4}$ into H$_{3}$S causes the dynamical stabilization of the high $T_{\rm c}$ phase at lower pressures and \ce{CSH_{7}} shows the $T_{\rm c}$ value of 181\,K at 100\,GPa~\cite{Sun2019_CHS}. Inspired by these theoretical results, experimentalists have explored novel ternary or multinary hydrides showing higher $T_{\rm c}$ at lower pressures. In 2020, Snider \textit{et al.} observed that carbonaceous sulfur hydride, i.e. the C-S-H system, shows a room-temperature superconductivity of $T_{\rm c} = 288$\,K at 267\,GPa~\cite{Snider2020}. However, in 2022, the article was retracted owing to a lot of questions about the data and results~\cite{Wang2021,Snider2022_retraction}. In 2022, Grockowiak \textit{et al.} reported ``hot superconductivity'' in ternary or multinary compounds based on lanthanum hydride~\cite{Grockowiak2022}. The authors claim that $T_{\rm c}$ is increased to 556\,K by subsequent thermal excursion to high temperatures, which might be induced by the reaction of La-H with other materials existing in the sample chamber of the diamond anvil cell (DAC). In 2023, Dasenbrock-Gammon \textit{et al.} observed a room-temperature superconductivity near ambient pressure, i.e. $T_{\rm c} = 294$\,K at 1\,GPa, in nitrogen-doped lutetium  hydride (Lu-N-H)~\cite{Dasenbrock-Gammon2023}, whereas the room-temperature superconductivity has not been reproduced by first-principles calculations~\cite{Sun2023} and other experiments~\cite{Ming2023}. 
Finally, the article was retracted as the request of the co-authors eight months later~\cite{Dasenbrock-Gammon2023_retraction}. 

In the experiments shown in Ref. \onlinecite{Dasenbrock-Gammon2023}, ammonia borane, \ce{NH_{3}BH_{3}}, was used as a hydrogen source material for the synthesis of La-H. 
Assuming that La-H is reacted with N included in the sample chamber of DAC, we theoretically searched for thermodynamically stable and metastable phases and superconducting phases in the La-N-H ternary system. In high pressure region above 100\,GPa, thermodynamically stable phases and superconducting phases in the La-N-H system have been explored using first-principles calculations and structure search algorithms\cite{Ge2021,DiCataldo2022}. However, high pressurization is a bottle neck in device applications of hydride superconductors, and the exploration of the stable and superconducting phases in low pressure region is crucial for the applications. Thus, we searched for the stable and superconducting phases at 20\,GPa in low pressure region using first-principles calculations and our originally developed algorithm for stable composition search, which is based on an evolutionary construction of a formation-enthalpy convex hull combined with universal neural network potential. In addition, considering the recent experimental observation of the room-temperature superconductivity near ambient pressure in N-doped Lu-H~\cite{Dasenbrock-Gammon2023}, we investigate the effect of N doping on the superconductivity in the La-H. 

\section{Computational details}
\label{computationaldetails}

\begin{figure}
\includegraphics[width=8.0cm]{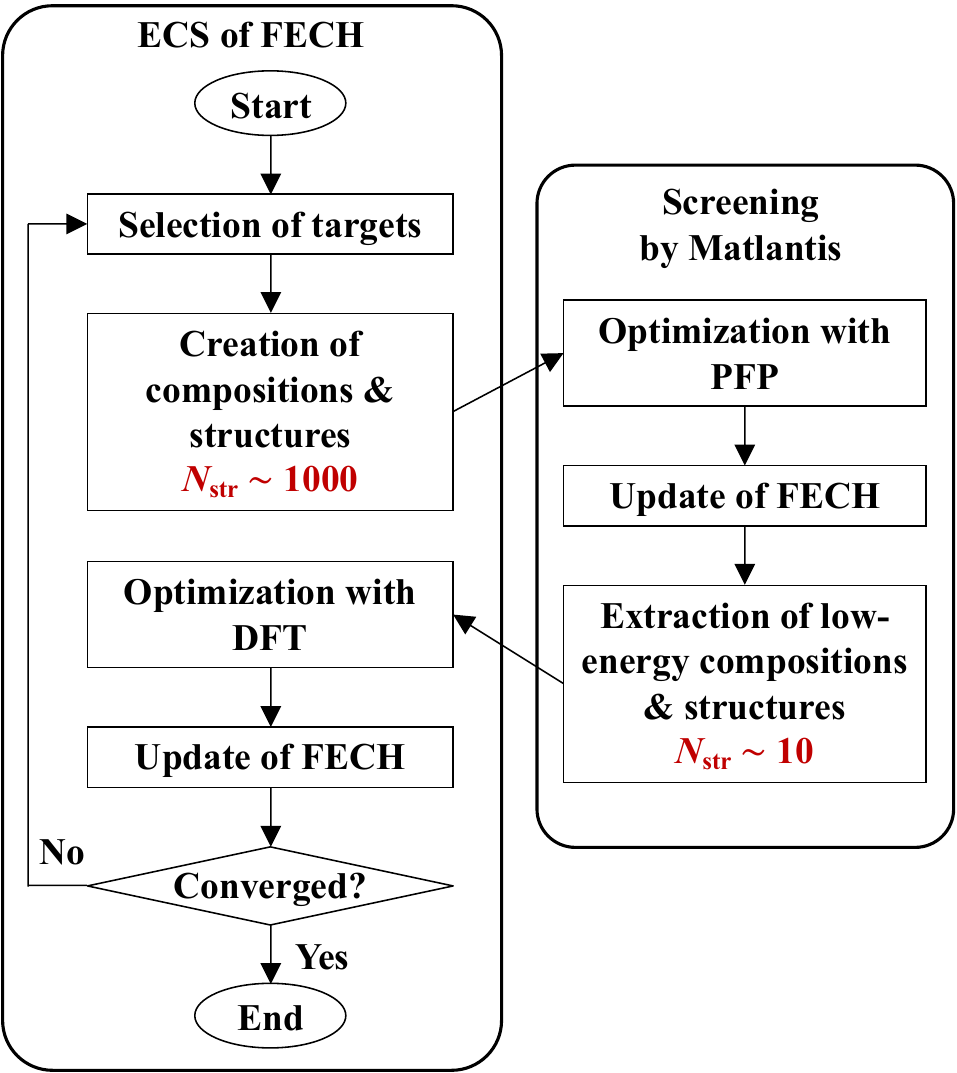}% Here is how to import EPS art
\caption{\label{Fig_flowchart} 
Flowchart of the evolutionary construction scheme (ECS) of a formation-energy convex hull (FECH) combined with density functional theory (DFT) calculations and the universal neural network potential (preferred potential, PFP) calculations. The PFP calculation is implemented in an atomic simulator, Matlantis. $N_{\text{str}}$ represents the number of crystal structures.}
\end{figure}
Since the search for thermodynamically stable phases in ternary systems is computationally expensive, 
we combined an evolutionary construction scheme (ECS) of a formation-enthalpy convex hull~\cite{Ishikawa2020-CH} with an atomic simulator using universal neural network potential, Matlantis~\cite{Matlantis}. In the ECS, compositions and structures with high probability of being stable are created by random applications of ``mating'', ``mutation'', and ``adaptive mutation'' to the compounds whose enthalpy difference ($\Delta H$) to the convex hull is small (see Figs. 2 and 3 in Ref. \onlinecite{Ishikawa2020-CH}). All the created structures are optimized and the convex hull is updated, which is defined as one generation. The stable and moderately metastable compounds are efficiently explored due to a data-driven approach based on the structural information of 
stable compounds~\cite{Ishikawa2021_Y-Co-B}, where ``moderately metastable'' means ``small $\Delta H$''. In the Matlantis, the potentials for more than 70 elements are being created using ten-million training data obtained by the density functional theory (DFT) calculations and a graph neural network and can universally be applied to a wide variety of systems including high-pressure conditions. The universal neural network potential is called ``preferred potential (PFP)''~\cite{Takamoto2022}. 
Figure \ref{Fig_flowchart} shows the flowchart of our algorithm. At each generation, a few thousand structures created by ECS are optimized using PFP and the convex hull is updated within Matlantis. Thanks to PFP, the optimization is rapidly and accurately accomplished with small calculation resources. 
The calculation speed and accuracy of Matlantis is compared with those of DFT in Table S1 and Figs. S1 and S2 in the Supplemental Material (SM)~\cite{SM_LaNH}.
Then, the candidates for thermodynamically stable and moderately metastable compounds are extracted based on $\Delta H$ for the convex hull diagram, where  the number of the structures are significantly decreased to a few tens. Then, only the extracted structures are optimized using the DFT calculations and the convex hull is updated. 

In our search, we used the Quantum ESPRESSO (QE) code~\cite{QE} for the DFT part. 
We set the upper limit of $\Delta H$ in ECS at 4.4\,mRy/atom for QE. 
We used the generalized gradient approximation by 
Perdew, Burke and Ernzerhof~\cite{PBE} for the exchange-correlation functional, and 
the Rabe-Rappe-Kaxiras-Joannopoulos ultrasoft pseudopotential~\cite{RRKJ90}. 
The energy cutoff was set at 80\,Ry for the wave function and 640\,Ry for the charge density. 
We adopted Marzari-Vanderbilt cold smearing of width 0.01\,Ry~\cite{M-V_smearing}.  
%The maximum number of atoms in calculation cell is 76, and 
%the $k$-space integration over the Brillouin zone (BZ) was carried out 
%on 24 $\times$ 24 $\times$ 24, 16 $\times$ 16 $\times$ 16, 12 $\times$ 12 $\times$ 12, 8 $\times$ 8 $\times$ 8, and 6 $\times$ 6 $\times$ 6 grids for the structures including 1--4, 5--12, 13--20, 21--30, and more than 30 atoms in the calculation cell, respectively. 
We increased the number of the $k$-point samplings in the Brillouin zone for the optimization until the formation enthalpy is sufficiently converged. 
We used v.3.0.0 for the version of PFP and the L-BFGS algorithm~\cite{LBFGS} for the optimization. 
We used $\Delta H \leq 11$\,mRy/atom as a condition for the candidate extraction by Matlantis. 
We performed the constant-pressure variable-cell optimization at 20\,GPa for the created structures. 

The superconducting $T_{\text{c}}$ was calculated using the Allen-Dynes modified McMillan formula~\cite{Allen-Dynes}, 
\begin{equation}
\label{AllenDynes}
T_{\text{c}}=\frac{f_{1}f_{2}\omega_{\log}}{1.2}
 \exp \left[ - \frac{ 1.04(1+\lambda )}
{ \lambda-\mu^{\ast}(1+0.62\lambda) } \right].
\end{equation}
The parameters, $\lambda$ and $\omega_{\log}$, are electron-phonon coupling constant and 
logarithmic-averaged phonon frequency, respectively, 
which represent a set of characters for the phonon-mediated superconductivity. 
$f_{1}$ and $f_{2}$ are correlation factors for the systems showing large $\lambda$. 
These parameters are defined as following, using Eliashberg function $\alpha^2F(\omega)$: 
\begin{eqnarray}
  \lambda
  &=& 2 \int_0^\infty
  d\omega\frac{\alpha^2 F(\omega)}{\omega},\\
  \omega_{\log} 
  &=&\exp \left[\frac{2}{\lambda}\int_0^\infty
  d\omega\frac{\alpha^2F(\omega)}{\omega}\log\omega\right], \\ 
  f_1
  &=&
  \left\{1+\left[\frac{\lambda}{2.46(1+3.8\mu^{\ast})} \right]^{3/2}\right\}^{1/3}, \\
  f_2
  &=&
  1+\frac{(\omega_2/\omega_{\log}-1)\lambda^2}
  {\lambda^2+[1.82(1+6.3\mu^{\ast})(\omega_2/\omega_{\log})]^2}, 
\end{eqnarray}
where 
\begin{eqnarray}
  \omega_2
  &=& \left[\frac{2}{\lambda}\int_0^\infty
  d\omega\alpha^2F(\omega)\omega\right]^{1/2}.
\end{eqnarray}
To obtain these parameters, we performed the phonon calculations implemented in the QE code. The effective screened Coulomb repulsion constant $\mu^{*}$ 
was assumed to be 0.10, which has been considered to be a reasonable value 
for hydrides. 
The $k$- and $q$-point grids for the calculations are listed in Table S2 in SM~\cite{SM_LaNH}.

\section{Results}

\begin{figure}
\includegraphics[width=8.6cm]{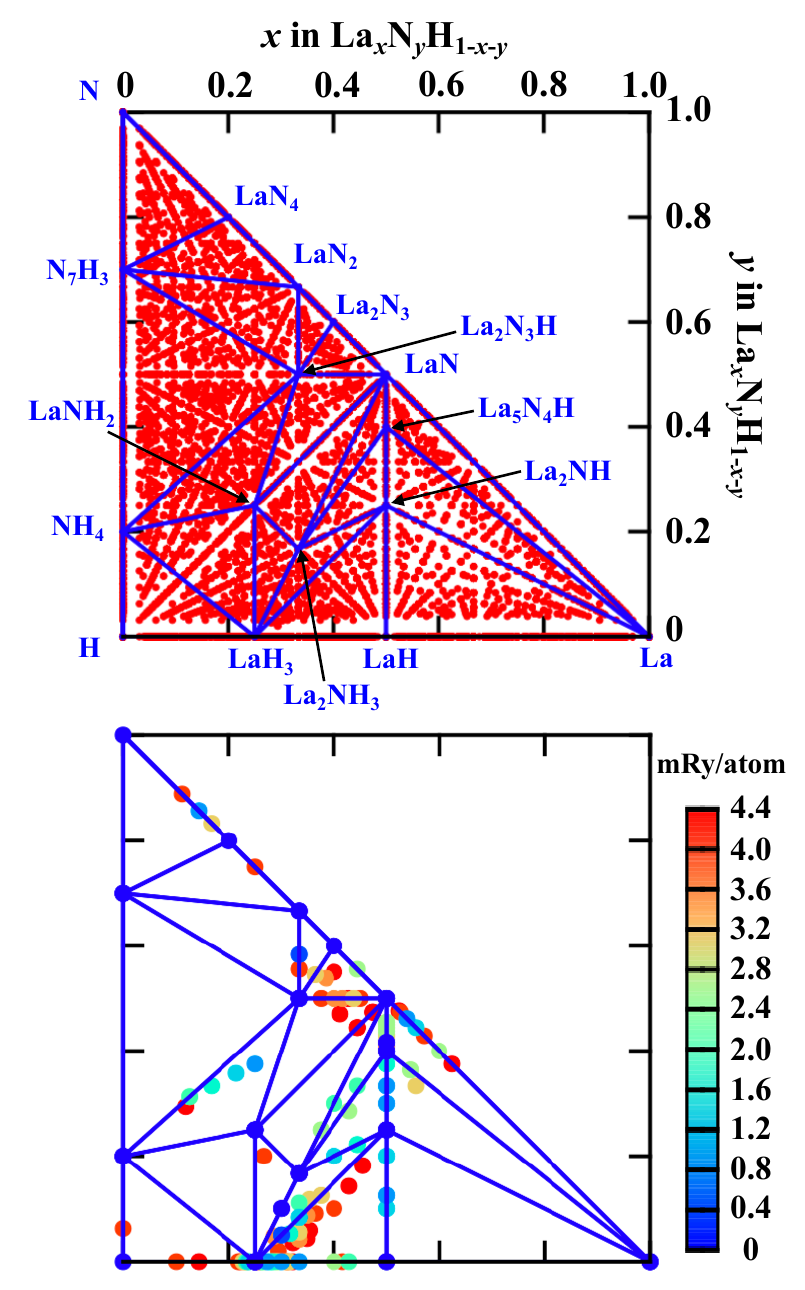}% Here is how to import EPS art
\caption{\label{Fig_convexhull_Matlantis} 
Formation-enthalpy convex hull diagram of the La-N-H system (\ce{La_{$x$}N_{$y$}H_{$1-x-y$}}) at 20\,GPa, obtained by Matlantis. The convex hull is projected on the $xy$ plane, and the vertices (the intersections of the lines) correspond to thermodynamically stable compounds. The upper panel shows all the compounds investigated in this study and the lower panel shows the moderately metastable compounds with $\Delta H$ less than 4.4\,mRy/atom.}
\end{figure}
Figure \ref{Fig_convexhull_Matlantis} shows the convex hull diagram of the formation enthalpy for the La-N-H system (\ce{La_{$x$}N_{$y$}H_{$1-x-y$}}, $0 \leq x \leq 1$, $0 \leq y \leq 1$) at the 20th generation obtained by Matlantis, at which the convex hull is sufficiently converged. In the upper panel, the convex hull is projected on the $xy$ plane, viewed along the $z$ axis showing the formation enthalpy. The lines and their intersections show the edges and vertices of the convex hull, respectively: the vertices correspond to thermodynamically stable compounds at 20\,GPa. The dots other than the intersections are compositions created by ECS. We obtained 13 thermodynamically stable compounds as follows: LaH$_3$ and LaH on the La-H line, NH$_4$ and N$_7$H$_3$ on the N-H line, LaN$_4$, LaN$_2$, La$_2$N$_3$, and LaN on the La-N line, and LaNH$_2$, La$_2$NH$_3$, La$_2$N$_3$H, La$_5$N$_4$H, and La$_2$NH in the triangle. In the lower panel, the moderately metastable compounds with $\Delta H$ less than 4.4\,mRy/atom are extracted. The compounds are concentrated on the line connecting between LaH and LaN, i.e. the line of $x = 0.5$, on the line connecting between LaN and La$_2$N$_3$H, and near the line connecting between La$_2$NH and LaH$_3$. 

\begin{figure}
\includegraphics[width=8.7cm]{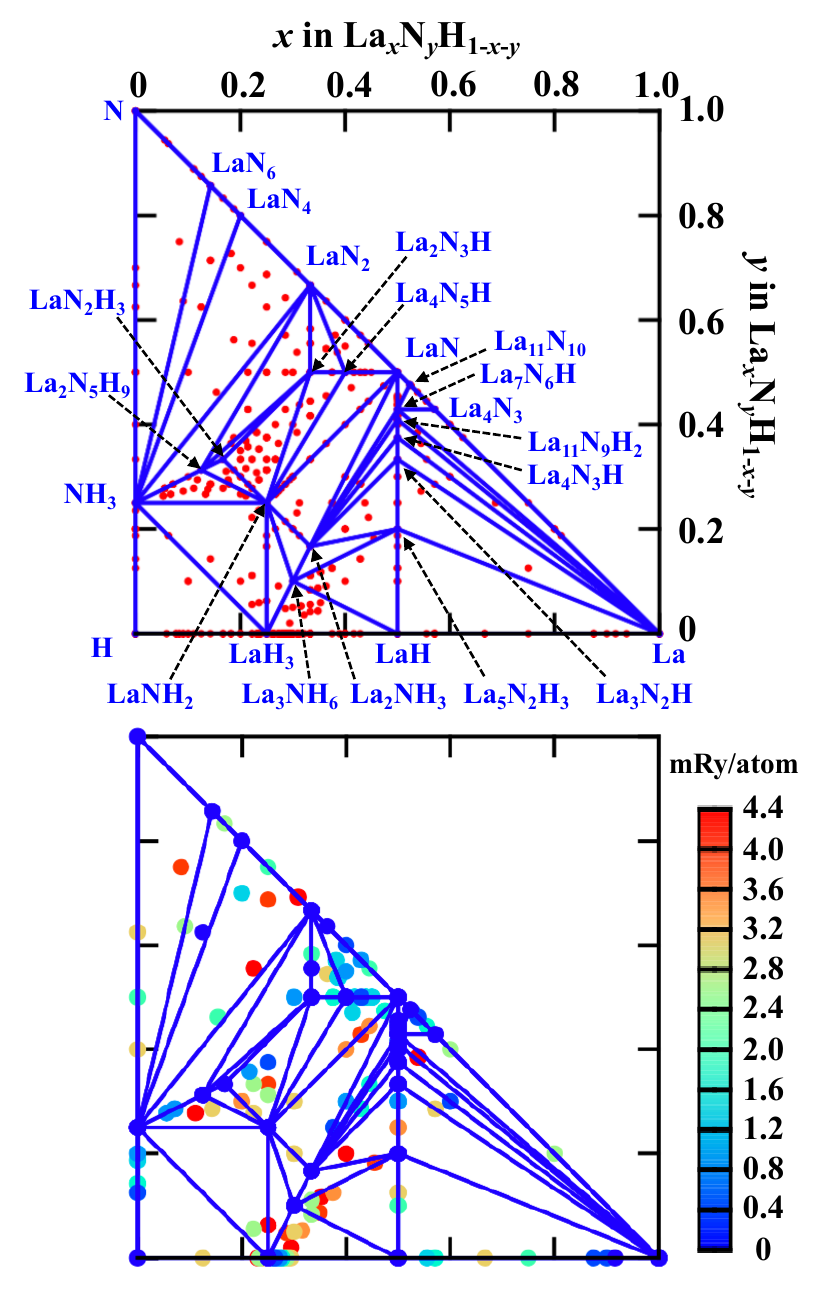}% Here is how to import EPS art
\caption{\label{Fig_convexhull_QE} 
Formation-enthalpy convex hull diagram of the La-N-H system (\ce{La_{$x$}N_{$y$}H_{$1-x-y$}}) at 20\,GPa, obtained by QE. The convex hull is projected on the $xy$ plane, and the vertices (the intersections of the lines) correspond to thermodynamically stable compounds. The upper panel shows all the compounds investigated in this study and the lower panel shows the moderately metastable compounds with $\Delta H$ less than 4.4\,mRy/atom.}
\end{figure}
Figure \ref{Fig_convexhull_QE} shows the similar convex hull diagram obtained by the QE optimization after the screening by Matlantis. The stable and metastable compounds are compared with those obtained by Matlantis in Table \ref{Comparison_Matlantis_QE}. Eight out of twenty six compounds were obtained as the stable compounds by both Matlantis and QE: LaH$_3$, LaH, LaN$_4$, LaN$_2$, LaN, LaNH$_2$, La$_2$NH$_3$, and La$_2$N$_3$H. Out of the other eighteen compounds, sixteen compounds are predicted to be stable by Matlantis or QE and moderately metastable ($\Delta H \le 4.4$\,mRy/atom) by the other. These results suggest that Matlantis has high predictability for thermodynamically stable compounds and is a competent precursor for QE. 
In contrast, Matlantis shows a low predictability for the compounds on the N-H line; NH$_3$ and N$_7$H$_3$ were predicted as quite unstable and stable compounds by Matlantis, respectively. See also two data largely deviated from the tendency of the other data in Fig. S2 in SM, which correspond to NH$_{4}$ and N$_7$H$_3$~\cite{SM_LaNH}. Thus, careful attention should be paid to the use of Matlantis for the N-H system under high pressure. 
\begin{table}
\caption{\label{Comparison_Matlantis_QE}
Comparison of thermodynamically stable and metastable phases between Matlantis and QE. The stable and moderately metastable ($\Delta H \le 4.4$\,mRy/atom) phases are indicated by circle and triangle, respectively. Unstable compounds appearing in the region of $\Delta H > 4.4$\,mRy/atom are indicated by a cross.}
%\begin{center}
\begin{ruledtabular}
\begin{tabular}{ccccc}
   System      &           &  Compounds   &  Matlantis  & QE \\
\hline
La-H & & LaH$_3$ & $\circ$ & $\circ$ \\
         & & LaH & $\circ$ & $\circ$ \\
N-H & & NH$_4$ & $\circ$ & $\triangle$ \\
       & & NH$_3$ & $\times$ & $\circ$  \\
       & & N$_7$H$_3$ & $\circ$ & $\times$  \\
La-N & & LaN$_6$ & $\triangle$ & $\circ$ \\
       & & LaN$_4$ & $\circ$ & $\circ$ \\
       & & LaN$_2$ & $\circ$ & $\circ$  \\
       & & La$_2$N$_3$ & $\circ$ & $\triangle$  \\
       & & LaN & $\circ$ & $\circ$  \\
       & & La$_{11}$N$_{10}$ & $\triangle$ & $\circ$  \\
       & & La$_{4}$N$_{3}$ & $\triangle$ & $\circ$  \\
La-N-H & $x = 0.5$ & La$_{5}$N$_{2}$H$_{3}$ & $\triangle$ & $\circ$ \\
       & & La$_{2}$NH & $\circ$ & $\triangle$ \\
       & & La$_{3}$N$_{2}$H & $\triangle$ & $\circ$ \\
       & & La$_{4}$N$_{3}$H & $\triangle$ & $\circ$ \\
       & & La$_{5}$N$_{4}$H & $\circ$ & $\triangle$ \\
       & & La$_{11}$N$_{9}$H$_{2}$ & $\triangle$ & $\circ$ \\
       & & La$_{7}$N$_{6}$H & $\triangle$ & $\circ$ \\
       & Others & LaNH$_{2}$ & $\circ$ & $\circ$ \\
       & & La$_2$NH$_{3}$ & $\circ$ & $\circ$ \\
       & & La$_2$N$_{3}$H & $\circ$ & $\circ$ \\
       & & La$_4$N$_{5}$H & $\triangle$ & $\circ$ \\
       & & LaN$_{2}$H$_{3}$ & $\triangle$ & $\circ$ \\
       & & La$_2$N$_{5}$H$_{9}$ & $\triangle$ & $\circ$ \\
       & & La$_3$NH$_{6}$ & $\triangle$ & $\circ$ \\
%H-La & LaH$_3$ & $\circ$ &  & $\circ$ & \\
\end{tabular}
%\end{center}
\end{ruledtabular}
\end{table}
\begin{figure}
\includegraphics[width=8.7cm]{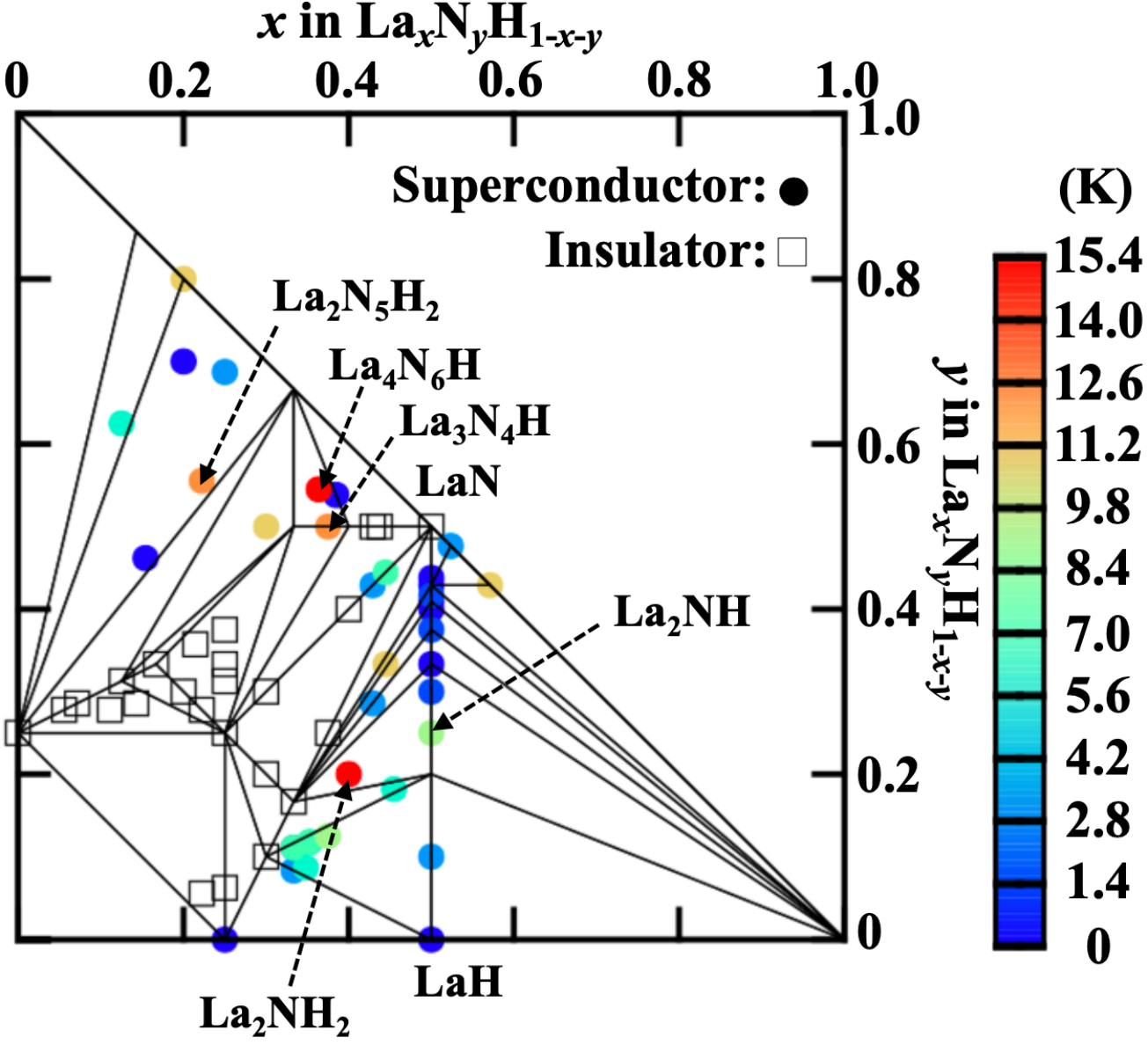}% Here is how to import EPS art
\caption{\label{Fig_Tc} 
Superconducting $T_{\rm c}$ data for the La-N-H system at 20\,GPa. The superconducting and insulating phases are indicated as closed circle and open square, respectively. The color of the circle represents the magnitude of the $T_{\rm c}$ value.}
\end{figure}
\begin{figure}
\includegraphics[width=8.7cm]{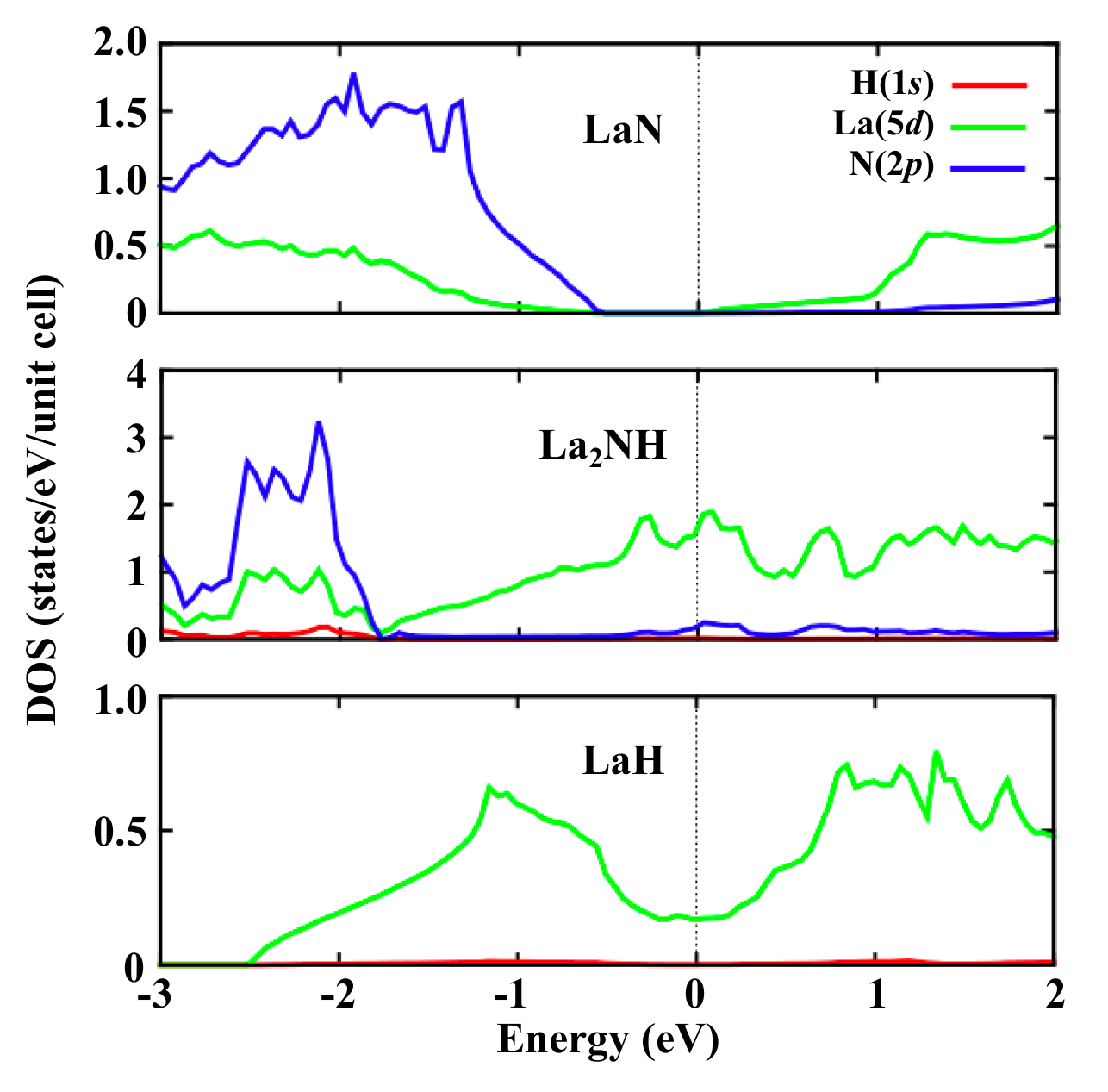}% Here is how to import EPS art
\caption{\label{Fig_DOS} 
Density of states (DOS) for LaH, La$_2$NH, and LaN at 20\,GPa. H($1s$), La($5d$), and N($2p$) represent the $1s$ states of H, the $5d$ states of La, and the $2p$ states of N, respectively. The Fermi level is set to zero. }
\end{figure}
\begin{figure}
\includegraphics[width=8.7cm]{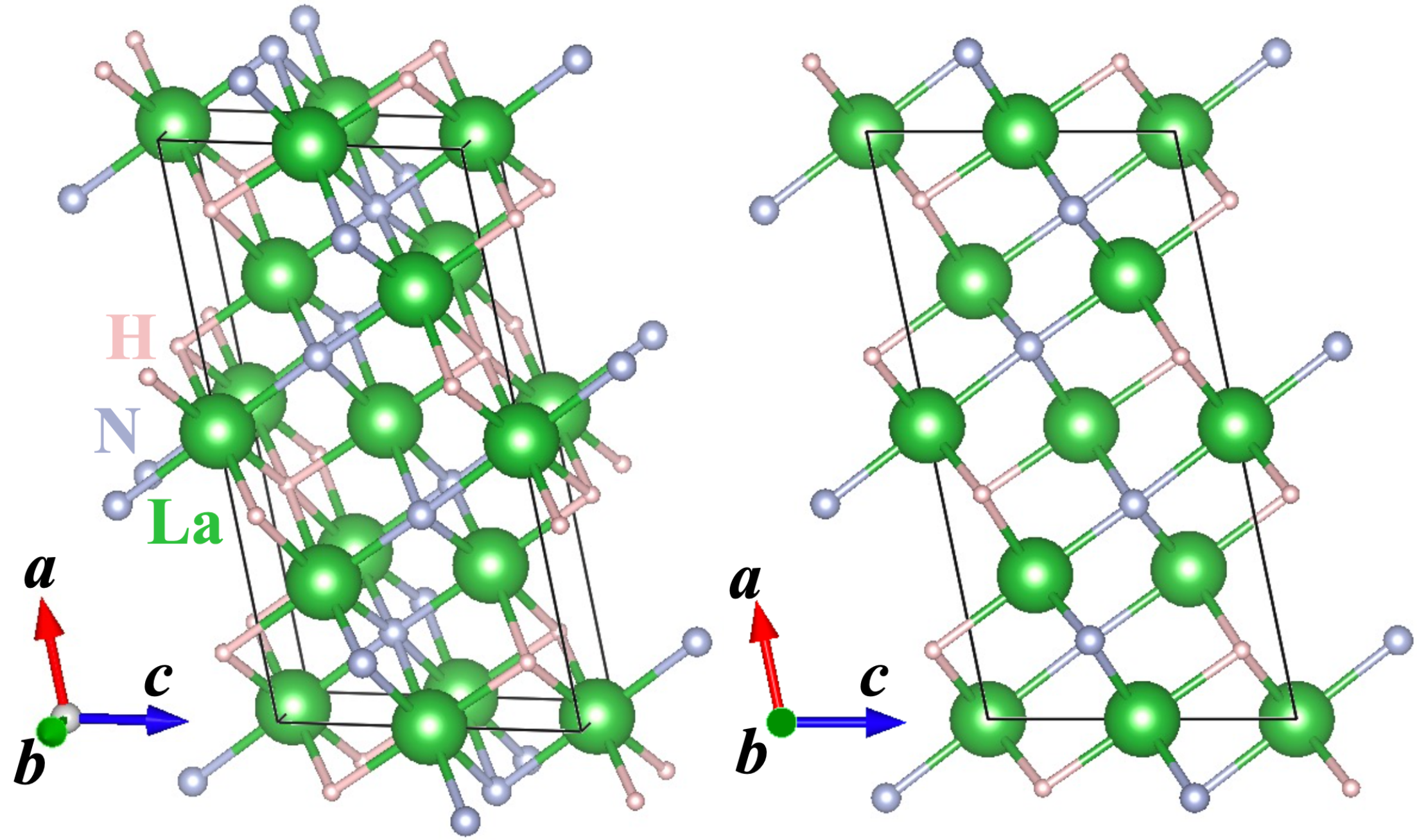}% Here is how to import EPS art
\caption{\label{Fig_La2NH} 
Crystal structure of La$_2$NH, assigned as a monoclinic $C2/m$. The right shows the structure viewed along the $b$ axis. The structure was drawn with VESTA~\cite{VESTA}.}
\end{figure}
\begin{table}
\caption{\label{Tcdata}
Superconductivity data of ternary compounds in the La-N-H system, predicted by the evolutionary construction scheme of a formation-enthalpy convex hull. $\Delta H$, $\lambda$, $\omega_{\log}$, and $T_{\rm c}$ are the enthalpy difference to the convex hull in the unit of mRy/atom, the electron-phonon coupling constant, 
the logarithmic-averaged phonon frequency in the unit of K, and the superconducting critical temperature in the unit of K, respectively.}
%\begin{center}
\begin{ruledtabular}
\begin{tabular}{ccccccc}
         & Compoud & Structure &  $\Delta H$ &  $\lambda$   &  $\omega_{\log}$  & $T_{\rm c}$ \\
\hline
$x = 0.5$ & La$_5$NH$_4$ & $C2/m$ & 2.12 & 0.5848 & 189.2 & 4.13 \\
                & La$_2$NH        & $C2/m$ & 3.51 & 0.7492 & 205.2 & 8.77 \\
                & La$_5$N$_3$H$_2$    & $C2/m$ & 1.05  & 0.4916 & 226.2 & 2.63 \\
                & La$_3$N$_2$H      & $P\bar{6}m2$ & 0 & 0.2802 & 279.6 & 0.07 \\
                & La$_4$N$_3$H      & $C2/m$ & 0 & 0.4127 & 280.1 & 1.42 \\
                & La$_5$N$_4$H      & $C2/m$ & 0.14 & 0.3852 & 300.0 & 1.03 \\
                & La$_{11}$N$_9$H$_2$     & $P\bar{1}$ & 0 & 0.3886 & 309.1  & 1.11 \\
                & La$_{6}$N$_{5}$H & $C2/m$ & 0.59 & 0.4614 & 294.2  & 2.60 \\
                & La$_7$N$_6$H     & $R\bar{3}$  & 0 & 0.3761 & 320.0  & 0.94 \\
                & La$_8$N$_7$H     & $P2/m$  & 0.33 & 0.3541 & 251.1 & 0.49 \\
Others     & La$_2$NH$_2$  & $P3m1$ & 4.12 & 0.9337 & 215.6 & 14.41 \\
                & La$_4$N$_6$H  & $P1$ & 2.99 & 0.9353 & 210.8 & 14.12 \\
                & La$_2$N$_5$H$_2$  & $Cm$ & 4.28 & 0.7592 & 274.2 & 12.08 \\
                & La$_3$N$_4$H  & $P1$ & 1.61 & 0.7492 & 280.0 & 11.95 \\
\end{tabular}
%\end{center}
\end{ruledtabular}
\end{table}
Next, we investigated the superconductivity of the La-N-H system following the convex hull diagram (Fig. \ref{Fig_Tc}). First we focus on the compounds existing on the line of $x = 0.5$. Figure \ref{Fig_DOS} shows the density of states (DOS) for LaH ($y = 0$), La$_2$NH ($y = 0.25$), and LaN ($y = 0.5$). LaH has the large dip at around the Fermi level in DOS and shows weak superconductivity: $\lambda$ is 0.2195 and $T_{\rm c}$ is less than 5\,mK. The DOS at the Fermi level gradually increases with the increase of N doping in LaH, and the $\lambda$ and $T_{\rm c}$ values increase to 0.5848 and 4.13\,K in La$_5$NH$_4$ ($y = 0.1$) and reach 0.7492 and 8.77\,K in La$_2$NH. The DOS data shows that the $5d$ states of La, which is defined as La($5d$) here, dominantly contribute to the electronic states at the Fermi level. La$_2$NH takes a monoclinic $C2/m$ structure, which is assigned using FYNDSYM~\cite{FINDSYM} (Fig. \ref{Fig_La2NH}). The superconductivity, however, is weakened by further N doping and the $T_{\rm c}$ values are less than 2.90\,K, and finally, LaN is an insulator with a band gap of 0.6\,eV. The superconductivity data and the space group of the structures are listed in Table \ref{Tcdata}. 

\begin{figure}
\includegraphics[width=8.7cm]{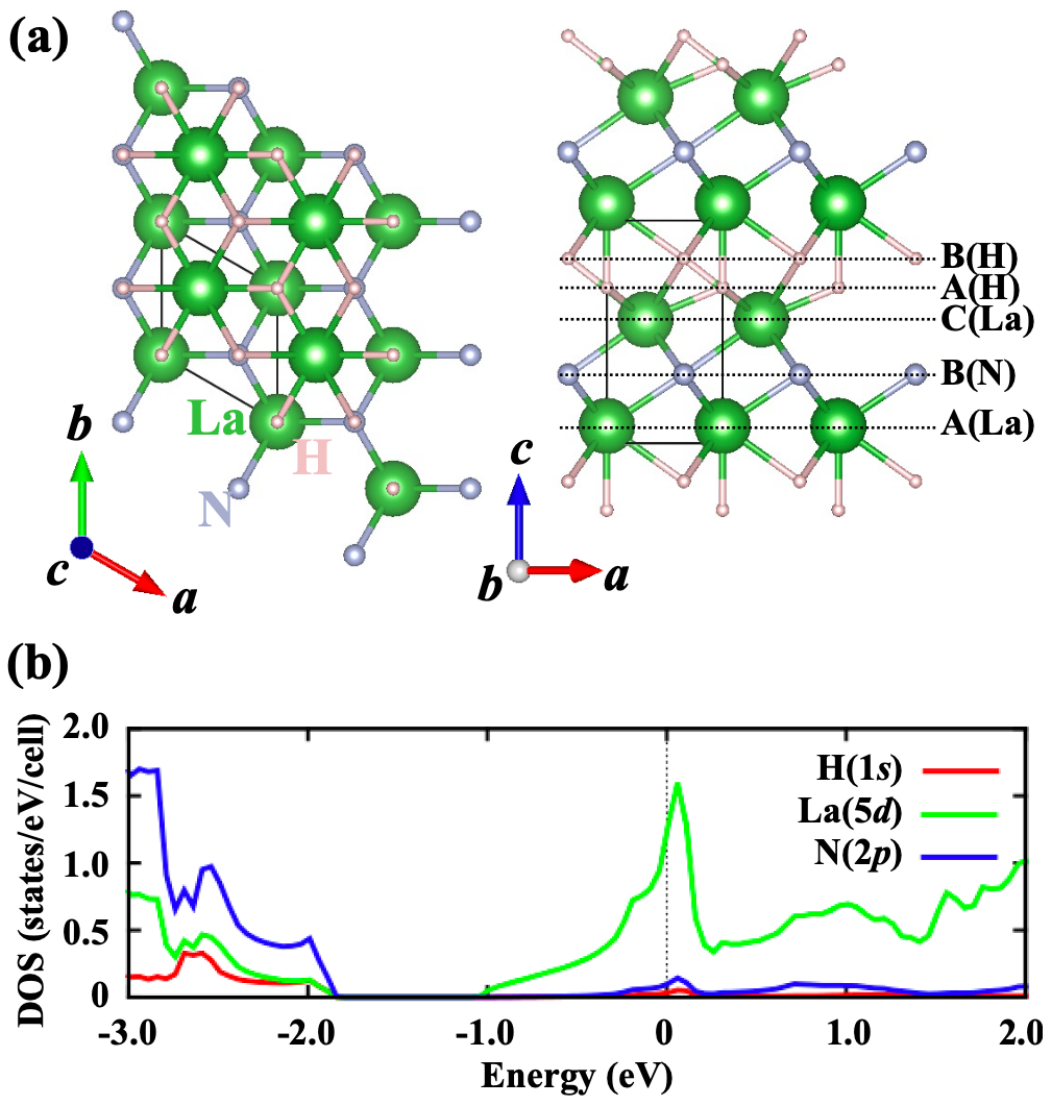}% Here is how to import EPS art
\caption{\label{Fig_La2NH2} 
(a) Crystal structure of La$_2$NH$_2$, assigned as a trigonal $P3m1$, viewed along the $c$ axis (left) and the $b$ axis (right). The stacking order A(La)-B(N)-C(La)-A(H)-B(H) is formed along the $c$ axis. The structure was drawn with VESTA~\cite{VESTA}. (b) Density of states (DOS) for La$_2$NH$_2$ at 20\,GPa. H($1s$), La($5d$), and N($2p$) represent the $1s$ states of H, the $5d$ states of La, and the $2p$ states of N, respectively. The Fermi level is set to zero.}
\end{figure}
Apart from the compounds on the line of $x = 0.5$, we explored other compounds showing higher $T_{\rm c}$ than La$_2$NH. We found that La$_2$NH$_2$ shows $\lambda = 0.9337$ and $T_{\rm c} = 14.41$\,K. La$_2$NH$_2$ takes a trigonal $P3m1$ structure. In this structure, the triangular lattice is formed in the plane parallel to the $ab$ plane and is stacked along the $c$ axis with the order of A(La)-B(N)-C(La)-A(H)-B(H), as shown in the right of Fig. \ref{Fig_La2NH2}a. Note that A, B, and C are the same as the layer positions describing the hexagonal close-packed structure (A-B) or the face-centered cubic structure (A-B-C). Figure \ref{Fig_La2NH2}b shows the DOS for La$_2$NH$_2$. The DOS has a large peak at around the Fermi level due to a dominant contribution of the $5d$ states of La, which results in the enhancement of the superconductivity. 
The top of the peak emerges at 0.1\,eV from the Fermi level, and the superconductivity is expected to be further enhanced by electron doping. 
We also predicted that La$_4$N$_6$H shows $\lambda = 0.9353$ and $T_{\rm c} = 14.12$\,K, which are comparable to the values of La$_2$NH$_2$. Unlike La$_2$NH$_2$, the $2p$ states of N dominantly contribute to the electronic states at the Fermi level.  
%We also predicted that La$_4$N$_6$H shows $\lambda = 0.9353$ and $T_{\rm c} = 14.12$\,K. This compound takes a triclinic $P1$ structure, which is formed by the doping of H atoms into metastable La$_2$N$_3$ (Fig. \ref{Fig_La4N6H}). 
%The doped H atoms are bonded with the N atoms occupying the $4b$ site in La$_2$N$_3$ with a tetragonal $I4_{1}/amd$. 
La$_2$N$_5$H$_2$ and La$_3$N$_4$H show the $T_{\rm c}$ values of 12.08\,K and 11.95\,K, respectively: $\lambda = 0.7592$ and $\omega_{\log} = 274.2$\,K for La$_2$N$_5$H$_2$ and $\lambda = 0.7492$ and $\omega_{\log} = 280.0$\,K for La$_3$N$_4$H. 
The $\lambda$ values are smaller than those of La$_2$NH$_2$ and La$_4$N$_6$H and are approximately same as that of La$_2$NH on the line of $x = 0.5$, 
whereas the $\omega_{\log}$ values are larger than those of the three compounds, which results in the $T_{\rm c}$ values of about 12\,K. In addition to these compounds, we predicted more than 20 compounds showing the superconductivity. Their superconductivity data are shown in Table S3 in SM~\cite{SM_LaNH}.
%
%\begin{figure}
%\includegraphics[width=8.7cm]{Fig_La4N6H.pdf}% Here is how to import EPS art
%\caption{\label{Fig_La4N6H} 
%(a) Crystal structure of metastable La$_2$N$_3$ with a tetragonal $I4_{1}/amd$, viewed along the $a$ axis. (b) Crystal structure of La$_4$N$_6$H with a triclinic $P1$ structure, viewed along the $a$ axis. The doped H atoms are bonded with the N atoms occupying the $4b$ site in $I4_{1}/amd$ La$_2$N$_3$.}
%\end{figure}
%

Although La$_4$NH$_{13}$ with $x = 0.2222$ and $y = 0.0556$ has the largest hydrogen concentration of all the stable and moderately metastable ternary compounds predicted by our search, this compound is the insulator with the band gap of 0.45\,eV. In addition, ternary compounds with $x \leq 0.3$ and $y \leq 0.4$ are all insulating phases (see Fig. \ref{Fig_Tc}), which are listed in Table S4 in SM~\cite{SM_LaNH}. These results suggest that superhydrides such as LaH$_{10}$ observed in a megabar region~\cite{Somayazulu2019,Drozdov2019-LaHx} are not expected to be obtained at 20\,GPa in the La-N-H ternary system. 

\section{Conclusions and discussion}

To predict novel hydride superconductors, we explored thermodynamically stable and metastable phases at 20\,GPa in the La-N-H ternary system (\ce{La_{$x$}N_{$y$}H_{$1-x-y$}}, $0 \leq x \leq 1$, $0 \leq y \leq 1$) integrating the evolutionary construction scheme of a formation-enthalpy convex hull, the universal neural network potential (PFP) calculations implemented in Matlantis, and the DFT calculations implemented in QE. Thanks to the screening by the PFP calculations, we quickly and accurately obtained stable and moderately metastable phases. The convex hull diagram shows that more than fifty compounds emerge as stable and moderately metastable phases in the region of $\Delta H \leq 4.4$\,mRy/atom, which can be obtained experimentally by high-pressure and high-temperature synthesis techniques. In particular, they are concentrated on the line connecting between LaH and LaN, i.e. the line of $x = 0.5$. Following the convex hull diagram, we investigated the superconductivity in the La-N-H system at 20\,GPa. For the compounds on the line of $x = 0.5$, we found that the superconductivity is gradually enhanced due to N doping for LaH and the $T_{\rm c}$ value reaches 8.77\,K in La$_2$NH with $y = 0.25$. In addition, we predicted that metastable La$_2$NH$_2$ shows the highest $T_{\rm c}$ value, 14.41\,K, of all the ternary compounds predicted in this study. Ternary compounds with $x \leq 0.3$ and $y \leq 0.4$ are all insulating phases. These results suggest that it is difficult to obtain high-$T_{\rm c}$ superconductivity in the La-N-H system at 20\,GPa. In addition, further investigations, e.g. changing the third element from N or changing pressure condition, are required to prove the hot superconductivity previously observed in the La-H based system by the electric resistivity measurements. 

% If you have acknowledgments, this puts in the proper section head.
\begin{acknowledgments}
This work was supported by JSR-UTokyo Collaboration Hub, CURIE, and JSPS KAKENHI under Grant-in-Aid for Scientific Research on Innovative Areas (Research in a proposed research area), HYDROGENOMICS: Creation of Innovative Materials, Devices, and Reaction Processes using Higher-Order Hydrogen Functions (21H00029), Scientific Research (C) (23K03316 and 20K03868), and Scientific Research (S) (20H05644). A part of the computation was performed using the facilities of the Supercomputer Center, the Institute for Solid State Physics, the University of Tokyo.
\end{acknowledgments}
% Create the reference section using BibTeX:
%\bibliography{References}

\begin{thebibliography}{44}
\bibitem[{SM_()}]{SM_LaNH}
\bibinfo{note}{See Supplemtal Material at [URL will be inserted by publisher]
  for calculation speed and accuracy of Matlantis and calculation conditions
  and superconductivity data for stable and moderately metastable compounds in
  the La-N-H system at 20{\,}GPa.}
\end{thebibliography}

\end{document}